\documentclass[letterpaper]{emulateapj} 
\usepackage{apjfonts}                   

\usepackage{epsfig,graphicx,latexsym,amsmath,amssymb}
\usepackage{natbib}
\usepackage{hyperref}
\usepackage{mathrsfs}
\usepackage{lastpage}

\bibpunct[,]{(}{)}{;}{a}{}{,}

\newcommand {\PN}{{\rm PN}}

\begin{document}

\author{
Pau Amaro-Seoane\altaffilmark{1,2}\thanks{e-mail: Pau.Amaro-Seoane@aei.mpg.de},
 Patrick Brem\altaffilmark{3}\thanks{email: pbrem@ari.uni-heidelberg.de},
 Jorge Cuadra\altaffilmark{4}\thanks{e-mail: jcuadra@astro.puc.cl}
\& Philip J. Armitage\altaffilmark{5}\thanks{e-mail: pja@jilau1.colorado.edu}
}

\altaffiltext{1}{Max Planck Institut f\"ur Gravitationsphysik
(Albert-Einstein-Institut), D-14476 Potsdam, Germany}
\altaffiltext{2}{Institut de Ci{\`e}ncies de l'Espai (CSIC-IEEC), Campus UAB,
Torre C-5, parells, $2^{\rm na}$ planta, ES-08193, Bellaterra,
Barcelona, Spain}
\altaffiltext{3}{Astronomisches Rechen-Institut, M{\"o}nchhofstra{\ss}e 12-14, 69120,
Zentrum f\"ur Astronomie, Universit\"at Heidelberg, Germany}
\altaffiltext{4}{Departamento de Astronom\'ia y Astrof\'isica, Pontificia Universidad Cat\'olica de Chile, Santiago, Chile}
\altaffiltext{5}{JILA, University of Colorado and NIST, at Boulder, 440 UCB, Boulder, CO 80309-0440, USA}

\date{\today}

\label{firstpage}

\title{The butterfly effect in the extreme-mass ratio inspiral problem}

\begin{abstract}
Measurements of gravitational waves from the inspiral of a stellar-mass compact
object into a massive black hole are unique probes to test General
Relativity (GR) and MBH properties, as well as the stellar distribution about
these holes in galactic nuclei.  Current data analysis techniques can provide
us with parameter estimation with very narrow errors.  However, an EMRI is not
a two-body problem, since other stellar bodies orbiting nearby will influence
the capture orbit. Any deviation from the isolated inspiral will
induce a small, though observable deviation from the idealised waveform which
could be misinterpreted as a failure of GR.  Based on conservative analysis of
mass segregation in a Milky Way like nucleus, we estimate that the possibility
that another star has a semi-major axis comparable to that of the EMRI is
non-negligible, although probably very small.  This star introduces an
observable perturbation in the orbit in the case in which we consider only loss
of energy via gravitational radiation.  When considering the two
first-order non-dissipative post-Newtonian contributions (the periapsis shift
of the orbit), the evolution of the orbital elements of the EMRI turns out to
be chaotic in nature.  The implications of this study are twofold. From the one
side, the application to testing GR and measuring MBHs parameters with the
detection of EMRIs in galactic nuclei with a millihertz mission will be even
more challenging than believed. From the other side, this behaviour could in
principle be used as a signature of mass segregation in galactic nuclei.
\end{abstract}

\maketitle

\section{Motivation}
\label{sec.motivation}

A stellar mass black hole or neutron star executes $\sim 10^{5-6}$ orbits during
the final year of inspiral toward a $\sim 10^6 \ M_\odot$ supermassive black
hole (MBH). The large number of cycles implies that a phase-coherent  measurement of
the inspiral, achievable through detection of low frequency gravitational
waves, would be a tremendously powerful probe of the spacetime near a black
hole \citep{pau07,hughes09}. Among other things, it would enable a precise
determination of the spin of the supermassive black hole, and a test of General
Relativity that is independent of current constraints derived from pulsar
timing data.

There is no foreseeable instrument sensitive enough to detect gravitational
waves from extreme mass ratio inspirals (EMRIs) over time scales comparable to
the orbital period. As a consequence, realizing the astrophysical and
gravitational physics promise of EMRIs requires an assurance that the inspiral
can be accurately modeled over many orbits using templates calculated by
solving the 2-body problem in General Relativity \citep[for a review, see
e.g.,][]{barack09}. It is therefore necessary to assess whether gas, stars
or other compact objects in the vicinity, could significantly perturb EMRI
trajectories.  In the case of gas, perturbations to stellar mass black holes or
neutron stars\footnote{White dwarf EMRIs are excluded here, because mass loss
from the compact object itself could form a dynamically significant disc even
if the background accretion flow is of low density\citep{zalamea10}.} are
securely negligible provided that accretion on to the black hole occurs in a
low density, radiatively inefficient flow \citep{narayan00}. Such flows are
much more common than dense accretion discs, which {\em would} yield observable
phase shifts during inspiral \citep{KocsisEtAl11}, at least at the relatively low
redshifts where EMRIs may be observed.

In this Letter, we quantify the nature and strength of possible perturbations
from point mass perturbers: low mass stars or compact objects in tight orbits
around the supermassive black hole. Any perturbers are unlikely to
orbit close enough to the EMRI to undergo strong interactions, so the regime of
interest is one where the third body is relatively distant and the interaction
weak. The Newtonian analog of this problem has been studied extensively in the 
context both of Solar System satellite evolution, and for transit timing 
variations of extrasolar planets \citep{dermott88,agol05,holman05,veras11}. 
In Newtonian gravity, perturbations are strong only at the location of 
mean motion resonances, and these have the effect of inducing small jumps in 
eccentricity upon divergent resonance crossing. This would already be interesting 
for the EMRI problem, since the jumps in eccentricity would result in a perturbation 
to the gravitational wave decay rate, and an eventual dephasing of the waveform.
However, as we will see, the inclusion of post-Newtonian corrections changes 
the evolution qualitatively. Computing trajectories that include the 
two first-order non-dissipative post-Newtonian corrections, we find evidence 
of dependence on initial conditions in the evolution of the perturbed inner binary, such that arbitrarily 
small variations in the initial orbit lead to significantly
different future behaviour.

\section{Astrophysical limits on perturbers}
Is it likely that a star or compact object will be present close enough to 
perturb the orbit of an EMRI? Excluding low mass MBHs 
($M_\bullet < 10^6 \ M_\odot$), where the stellar tidal disruption limit 
comes into play, the existence of perturbers is not excluded by elementary 
arguments. Neither, however, is it easy to calculate the probability 
distribution of perturbers, whose proximity will depend upon the details 
of discreteness and relativistic effects very close to the 
MBH, and mass segregation and EMRI injection mechanisms in galactic 
nuclei \citep{PretoMerrittSpurzem04,FAK06a,ASEtAl04}.

Rather than face these difficulties, we limit ourselves here to order of 
magnitude estimates for the likely location of the nearest star and compact 
object. For stars, assumed to be of a single mass $M_*$, we assume a cusp-like 
distribution with density profile $\rho \propto R^{-\gamma}$, extending from the 
MBH to its radius of influence $R_{\rm BH} = GM_\bullet / \sigma^2$. 
Here $\sigma$ is the velocity dispersion of the galaxy. Using the fact that the enclosed mass, 
$M(R) \simeq M_\bullet$ at $R = R_{\rm BH}$, we find that the expected radius of the 
innermost star, $R_1$, is,
\begin{equation}
 \frac{R_1}{R_g} = \left( \frac{M_*}{M_\bullet} \right)^{1/(3-\gamma)} \left( \frac{c}{\sigma} \right)^2,
\end{equation}
where $R_g = GM_\bullet / c^2$. This formula yields an explicit estimate for $R_1$ 
once we adopt a relation between $M_\bullet$ and $\sigma$ \citep{gultekin09}. For the 
location of the next nearest compact object (or EMRI), we use an even simpler approach. We 
calculate the expected semi-major axis for uncorrelated inspirals due to gravitational 
radiation \citep{peters64}, assuming near-circular orbits and rate $\dot{N}_{\rm EMRI}$. 
Finally, we plot the tidal limit \citep[e.g.][]{rees88} for $0.3 \ M_\odot$ main-sequence 
stars.

\begin{figure}
\includegraphics[width=\columnwidth]{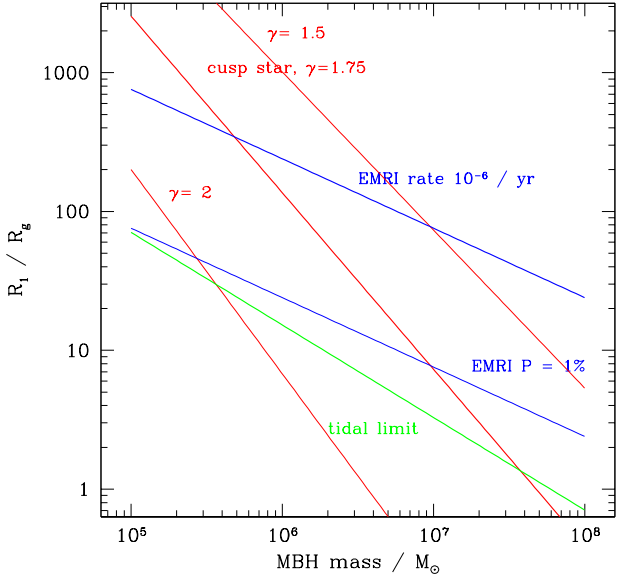}
\caption
   {Estimates for the semi-major axis of the innermost perturbing body around a 
   massive black hole, scaled to the hole's gravitational radius $R_g = GM_\bullet / c^2$.  
   The red lines show the location of the innermost star, estimated assuming that 
   stars of mass $0.3 \ M_\odot$ follow a single power-law cusp of index $\gamma$ in  
   a galaxy on the $M_\bullet$-$\sigma$ relation. The green line shows the tidal 
   disruption limit for such stars. The blue lines show the average (upper) 
   and 1\% probability (lower) location of the next nearest EMRI, assuming uncorrelated 
   inspirals at a rate of $10^{-6} \ {\rm yr}^{-1}$.}
\label{fig.R_1}
\end{figure}

Figure~\ref{fig.R_1} shows these estimates as a function of $M_\bullet$. For a standard 
cusp slope $\gamma = 1.75$, there is likely to be a low mass stellar perturber within 
a few hundred $R_g$ for $M_\bullet > 10^6 \ M_\odot$. Similarly, if the EMRI rate is 
as high as $10^{-6} \ {\rm yr}^{-1}$, there is a significant chance (at least a few 
percent) that a second compact object will be present between $10 - 10^2 \ R_g$ for 
$10^6 \ M_\odot < M_\bullet < 10^7 \ M_\odot$. Clearly, these crude estimates do not 
demonstrate that {\em most} EMRIs will be perturbed by third bodies, but they do 
suggest that perturbers may be close enough in some galaxies to motivate detailed 
consideration of their dynamical effects.

\section{Methods}

We are interested in the secular effect of a star acting on an EMRI which
will describe thousands of orbits in the detector bandwidth and
slowly decay. The kind of effects on the wave that we are looking at are tiny,
though detectable, and the mass difference between the two binaries (the
MBH-EMRI and the MBH-star systems) is huge. We need therefore a numerical tool
capable of integrating the plunging orbit of the EMRI while inducing a minimal
error in the integration, since data analysis techniques can detect e.g.
eccentricity differences of the order $\Delta e \sim 10^{-3}$
\citep{AS10a,PorterSesana10,KeyCornish11}. We hence have chosen to use a direct
$N-$body approach~\citep{Aarseth99,Aarseth03}, the {\tt planet} code, written
by Aarseth\footnote{who, as is his admirable custom, has made the code publicly
available \url{http://www.ast.cam.ac.uk/~sverre/web/pages/nbody.htm}}.
This is the most expensive method because it involves integrating all
gravitational forces for all three bodies at every time step, without making
any a priori assumptions about the system. Our approach employs the improved
Hermite integration scheme, which requires computation of not only the
accelerations but also their time derivatives.  Since we are simply integrating
Newton's equations directly, all gravitational effects are included. For the
purpose of our study, nonetheless, we have included relativistic corrections to
the Newtonian forces (the forces can be found in the same page in the {\tt toy}
code\footnote{\url{ftp://ftp.ast.cam.ac.uk/pub/sverre/toy/README}}). This was
first implemented in a direct-summation $N-$body code by \cite{KupiEtAl06}.
For this, one has to add perturbations in the integration, so that the forces
are modified by

\begin{equation}
{F}  = \overbrace{{F}_0}^{\rm Newt.}
+\overbrace{\underbrace{c^{-2}{F}_2}_{1\PN} +
\underbrace{c^{-4}{F}_4}_{2\PN}}^{\rm periapsis~shift} +
\overbrace{\underbrace{c^{-5}{F}_5}_{2.5\PN}}^{\rm energy~loss} +
\overbrace{\mathcal{O}(c^{-6})}^{\rm neglected}
\label{eq.F_PN}
\end{equation}

\noindent
In the last equation ``PN'' stands for post-Newtonian.
We note that the perturbations do not need to be small compared to the two-body
force \citep{Mikkola97}. 
The expressions for $F_2$, $F_4$ and $F_5$ can be found in \cite{BlanchetFaye01}, their
equation 7.16.

\section{Dissipation of energy and resonances}

We first analyse the system by contemplating only the relativistic effect of
dissipation of energy; i.e. our simulations only incorporate the 2.5 PN
correction term.  We stop the integration when the separation between the
stellar BH and the MBH is $a_{\bullet} = 5\,R_{\rm Schw}$, which approximately
corresponds to the limit where the PN approximation is not valid anymore. The
inspiral down to this distance takes typically in our simulations some 440,000
orbits.

\begin{figure}
\resizebox{0.8\hsize}{!}
          {\includegraphics[scale=1,clip]{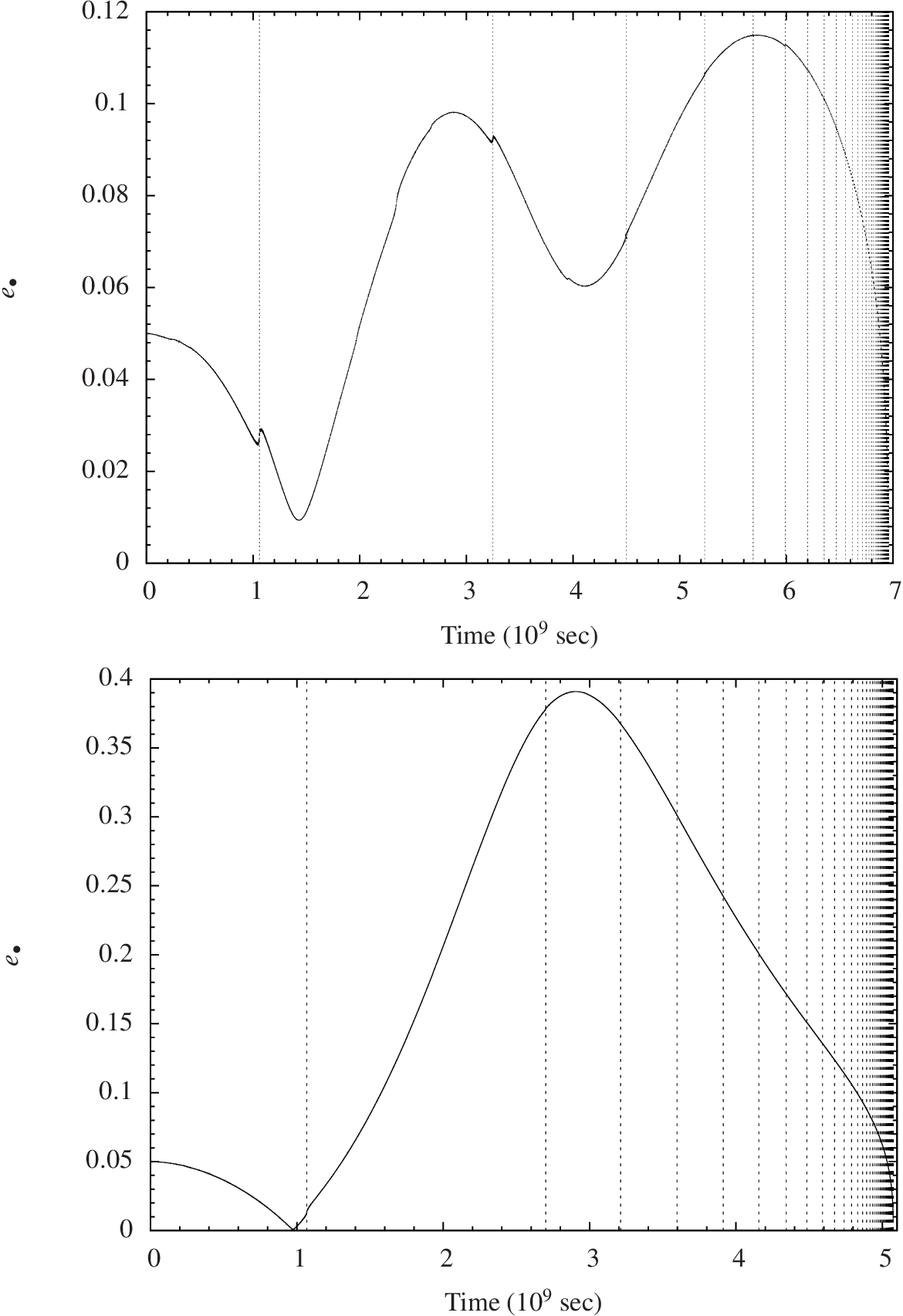}}
\caption
   {
{\em Upper panel:}
Results for the fiducial case using the direct-summation $N-$body integrator.
The mass of the MBH is ${\cal M}_{\bullet} = 10^{6}\,M_{\odot}$, the mass of
the stellar black hole is $m_{\bullet} = 10\,M_{\odot}$. See text for more details.
{\em Lower panel:} Same configuration but with an initial inclination of
the star of $i_{\star}=45^{\circ}$ instead of $30^{\circ}$, i.e. the
inclination triggers the Kozai mechanism, since $i_{\star} > 39.2^{\circ}$ {\em
and} the orbit is prograde.  As mentioned in the previous case, even if the
changes in eccentricity cannot be directly seen in the curve, they are of the
order $\Delta\,e_{\bullet} \sim 10^{-3}$.
   }
\label{fig.25PN}
\end{figure}

In Fig.(\ref{fig.25PN}) the test stellar black hole of mass $m_{\bullet} =
10\,M_{\odot}$ has been initially set in such an orbit that it is totally
embedded in a LISA-like detector band (i.e. with an orbital period $< 10^5$
secs, namely $P_{\bullet} = 6\times 10^3$ secs) and is hence an EMRI; its
initial semi-major axis is $a_{\bullet,\,\rm i} \simeq 1.45\times 10^{-6}$ pc
and its eccentricity $e_{\bullet,\,\rm i} = 0.05$. The perturber, a star of
mass $m_{\star} = 10\,M_{\odot}$ is initially on an orbit in which the
semi-major axis has the value $a_{\star ,\,\rm i} \simeq 4.1 \times 10^{-6}$ pc
and the eccentricity at $T=0$ is $e_{\star,\,\rm i} = 0.5$.  The inclination of
the system EMRI -- star was set to $30^{\circ}$ initially in the upper panel.
This constitutes our reference system. 

In the figure, the straight lines mark the condition $P_{\star}/P_{\bullet} =
A$, with $A$ an integer, $P_{\star}$ the period of the star around the MBH and
$P_{\bullet}$ the period of the EMRI around the MBH; i.e. where the resonances
occur. The first three resonances have an impact on $e_{\bullet}$ which can be
seen on the plot; later resonances do also affect $e_{\bullet}$, with
$\Delta\,e_{\bullet} \sim 10^{-3}$. We also note that in the upper panel one
can see in-between smaller jumps; they correspond to higher-order resonances,
$P_{\star}/P_{\bullet}$ = 5.5, 6.5 and 7.5.

We made the choice for an initial inclination of $30^{\circ}$ to avoid another
effect that introduces a change in the eccentricity.  In the lower panel we
have {\em exactly} the same system but for $i_{\star} = 45^{\circ}$. With this
value, and the fact that the orbit is prograde, the Kozai oscillation of
eccentricity is present \citep{Kozai62}. Even if the eccentricity of the EMRI
$e_{\bullet}$ suffers the characteristic Kozai oscillations, the loci for the
resonances still fulfill the condition $P_{\star}/P_{\bullet} ={\rm integer}$.

\section{Does the flap of the star at apoapsis set off a tornado at periapsis?}

\begin{figure}
\resizebox{\hsize}{!}
          {\includegraphics[scale=1,clip]{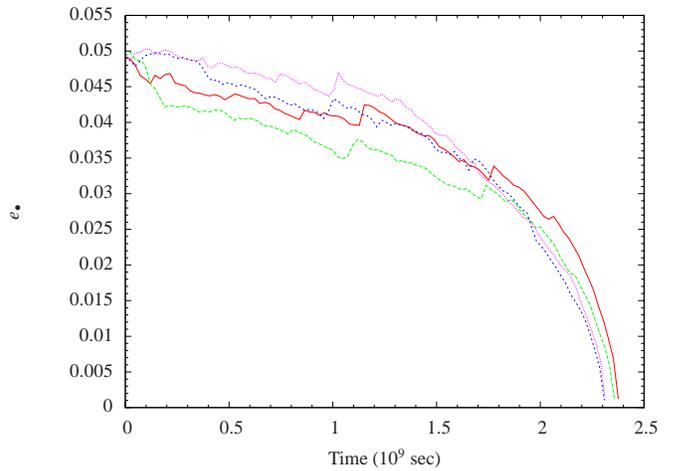}}
\caption
   {  
Fiducial case with energy dissipation and periapsis shift correcting terms for
different initial inclinations of the perturber. The solid (red) curve
corresponds to $i_{\star}=30^{\circ}$, the long-dashed (green) to
$i_{\star}=30.001^{\circ}$, the short-dashed (blue) corresponds to the fiducial
case plus a {\em billionth} of a degree, $i_{\star}=30.0000000001^{\circ}$ and
the dotted (magenta) to the reference plus a $10^{-13}$ of a degree,
$i_{\star}=30.0000000000001^{\circ}$.
   }
\label{fig.Fiducial_1225PN_30_30billionth}
\end{figure}

In this subsection we address numerically the effect of including the
relativistic periapsis shift along with the dissipation of energy; i.e. the set
of corrections as specified in Eq.(\ref{eq.F_PN}).  As we show below, the
effect of the periapsis shift changes completely the evolution of the system.
In Fig.(\ref{fig.Fiducial_1225PN_30_30billionth}) we show four cases. One of
them corresponds to the reference system but taking into
account the periapsis shift. 
We only display these examples but note that the behaviour is
also chaotic\footnote{When we use the word, we do not follow the rigorous
mathematical definition of chaos. We mean a strong
dependence on the initial conditions.} for other nearby choices of $i_{\star}$.
When using an initial inclination of $i_{\star} = 45^{\circ}$, which
corresponds to the same situation as in the lower panel of Fig.(\ref{fig.25PN})
but taking into account the periapsis shift, along with another case which is
identical but for $i_{\star} = 45.0000000001^{\circ}$, we find 
also a chaotic result which moreover eliminates the secular Kozai oscillation of $e$.

We have systematically studied this chaotic behaviour by running hundreds of
simulations in which we methodically increase in minimal differences an initial
dynamical orbital parameter such as the inclination, semi-major axis or
eccentricity. In all cases and parameters the evolution corroborates the
chaotic behaviour of the system. We have also tested a mass for the perturbing
star of $5$ and $1.44\,M_{\odot}$, as well as different values of $e_{\star}$
(0.1, 0.3, 0.7 and 0.9), with similar results.

\begin{figure}
\resizebox{\hsize}{!}
          {\includegraphics[scale=1,clip]{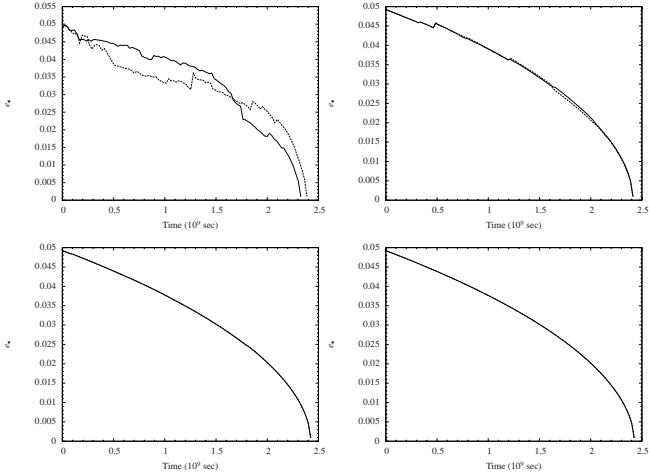}}
\caption
   {  
Same as in Fig.(\ref{fig.Fiducial_1225PN_30_30billionth}) but we set initially
the perturber at a larger and larger initial semi-major axis. From the top to
bottom and from the left to the right, the semi-major axis of the perturber is
$a_{\star} = 4\times10^{-6}$ pc, $6\times10^{-6}$ pc, $9\times10^{-6}$ pc and
$4.07243\times10^{-5}$ pc. Solid lines correspond to $i_{\star} = 30^{\circ}$
and dashed lines to $i_{\star} = 30.0000000001^{\circ}$.
   }
\label{fig.Fiducial_1225PN_30_30billionth_Different_Semimajor}
\end{figure}

In order to fence in the region within which the system is chaotic, we
systematically increase the semi-major axis of the star and run the same
experiment.  We start with the same difference in inclination at a slightly
larger semi-major axis, and then regularly increase it until we reach one order
of magnitude over the fiducial case, as we depict in
Fig.(\ref{fig.Fiducial_1225PN_30_30billionth_Different_Semimajor}). The chaotic
behaviour ceases at about one order of magnitude of the initial value of
$a_{\star}$ in the reference case.

\section{Quantifying the dependence on initial conditions of the system}

In this section we present a way of characterizing the rate of separation of
infinitesimally close trajectories systematically.  To achieve this we compare
our fiducial model with another case in which we set up the EMRI in an (almost)
imperceptibly different initial orbit (the initial difference is $2\times
10^{-10}$ pc, while the objects are moving on the same ellipse) and keep the
same initial conditions of the MBH and the perturber. Hence EMRI in the second
case differs only from the reference case slightly and has an initial distance
separation of $r_0$. We say that the two models are in phase provided that

\begin{equation}
r \approx r_0.
\label{eq:inphase}
\end{equation}

\begin{figure}
\resizebox{0.8\hsize}{!}
          {\includegraphics[scale=1,clip]{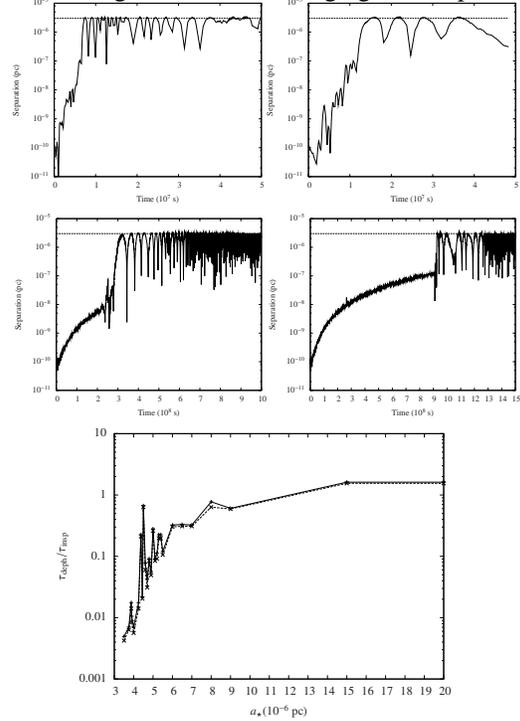}}
\caption
   {  
{\em Upper panels:}   
From the left to the right and from the top to the bottom we show the
separation $r$ for a increasing separation of the perturbing star of $3.5\times
10^{-6},\,3.9\times 10^{-6},\, 4.375\times 10^{-6},\,{\rm and~} 4.5\times
10^{-6}$ pc.  The dashed line shows the critical distance $2\,a_{\bullet}$.
Note the different timescales in the lower panels.
{\em Lower panel:}
$\tau_{\rm deph}$ against distance to the perturber normalized to the
gravitational radiation timescale of the isolated system $\tau_{\rm insp}$; i.e.
the merger timescale without the perturber acting onto the binary MBH-EMRI.
   }
\label{fig.MovingPertFindingLimitsChaos}
\end{figure}

\noindent
If the two different realizations reach a separation

\begin{equation}
 r \approx 2 \, a_{\bullet}
\label{eq:dephase}
\end{equation}

\noindent
the EMRI bodies are moving out of phase, on entirely unrelated orbits.
We thence are able to estimate a characteristic timescale ${\tau}_{\rm deph}$
for the system to become out of phase.
In Fig.(\ref{fig.MovingPertFindingLimitsChaos}) we display the separation
of the two systems for different distances to the perturber. From these
figures we can measure the value of a characteristic timescale 
$\tau_{\rm deph}$ for a given $a_{\star}$.

From the data points obtained in the upper panels of
Fig.(\ref{fig.MovingPertFindingLimitsChaos}) we
can then derive the relation displayed in the lower panel. For large
enough distances, of the order of $\sim 10^{-5}$ pc the two timescales converge
and the system becomes deterministic.

\section{Conclusions}

In this paper we have addressed the role of a perturbation on an EMRI by a
nearby star. The system depends extremely on {\em minimal} changes in the initial
conditions (as small as a $10^{-9}$ part in the inclination) lead to a very
different dynamical evolution. In all cases, however, the Kozai mechanism is
washed out by the periapsis shift, as one can expect \citep[see
e.g.][]{HolmanEtAl97,BlaesEtAl02}.  For distances of the order of $a_{\star}
\sim 10^{-5}$ pc the system enters the chaotic regime, for perturbing masses as
small as $1.44\,M_{\odot}$.  While we cannot state clearly whether this will be
a common feature for EMRIs, since the different dynamical and relativistic
phenomena involved in the problem are many and not straightforward (see for a
review \citealt{pau07} and also \citealt{pau11} for a dedicated review of the
dynamics), it seems plausible that for a Milky Way-like galaxy a star can be at
such a radius from the EMRI system that it will significantly perturb it. From
the standpoint of detection and data analysis, this is yet another complication
of the problem and could even lead to the misinterpretation that nature's GR is
not what we believe it to be.  On the other hand, from the point of view of
stellar dynamics, the detection of one of these systems would shed light on our
current understanding of galactic dynamics in general and mass segregation in
particular.

\acknowledgments 

We thank Marc Freitag and Rainer Sch{\"o}del for comments on the manuscript.
PJA acknowledges support from the NSF (AST-0807471), from NASA's Origins of
Solar Systems program (NNX09AB90G), and from NASA's Astrophysics Theory program
(NNX11AE12G).  PAS and JC were supported in part by the National Science
Foundation under Grant No. 1066293 and thank the hospitality of the Aspen
Center for Physics.  JC acknowledges support from FONDAP (15010003), FONDECYT
(11100240), Basal (PFB0609) and VRI-PUC (Inicio 16/2010), and the hospitality
of JILA, AEI and MPE.  PB and PAS acknowledge financial support for research
visits in China by The Silk Road Project (2009S1-5) of Chinese Academy of
Sciences, National Astronomical Observatories of China and PB also the
University of Heidelberg for travel costs through the excellence initiative ZUK
49/1, TP 14.8 International Research.  It is a pleasure for PAS to thank Sabine
Pendl for her extraordinary support at the Rote Insel during the preparation of
this work and also for her interest in non-linear dynamics and complex systems,
which motivated very interesting discussions.

\label{lastpage}
\end{document}